# Desalination Performance of Nano porous MoS2 Membrane on Different Salts of Saline Water: A Molecular Dynamics Study

Nudrat Nawal*, Md Rashed Nizam*, Priom Das, A K M Monjur Morshed

*Department of Mechanical Engineering, Bangladesh University of Engineering & Technology, Dhaka1000, Bangladesh*

**Joint first author*

## Abstract

The freshwater crisis is a growing concern and a pressing problem for the world because of the increasing population, civilization, and rapid industrial growth. Climate change causing droughts, the shrinking of glaciers, and the rising of sea levels are contributing to the crisis of fresh water. On average, one in nine people faces a lack of pure drinking water worldwide. The water treatment facilities are able to supply less than 1% of the total water demand. Water desalination can be a potential solution to deal with this alarming issue. Researchers have been exploring for quite some time to find novel nano-enhanced membranes and manufacturing techniques to increase the efficiency of the desalination process. Different forms of nano porous carbon nanostructures, for example, graphene sheets, carbon nanotubes, and multilayer nano porous graphene layers have shown astounding performance in the sector of water desalination. Graphene and graphene modified membranes showed huge potential as desalination membranes for comparatively easier synthesis process and higher ion rejection rate than conventional filter materials. Currently, single-layer $MoS_2$ has been discovered to have the same potential of water permeability and ion rejection rate as graphene membrane in a more energy-efficient way. For almost analogous nano porous structure of the graphene membrane, almost 70% of the higher water flux is obtained from the $MoS_2$ membrane. This has opened the way to analyze the performance of $MoS_2$ membrane furthermore. In this work, by using a molecular dynamics approach, it has been shown that nano porous $MoS_2$ membranes provide a promising result for desalinating other salts of seawater alongside NaCl. We have also observed the effect of variations in ions, pore size, and pressure on water permeation and ion rejection rates in the water desalination process. Water permeation and ion rejection rates are two important factors in examining the performance of a nano porous membrane for water desalination. In this study, water permeation increased significantly by increasing the pore area from 20Å to 80Å. The rate of water filtration increases in proportion to both applied pressure and pore size, sacrificing the ion rejection rate for the type of ions studied. A combination of salt ions in the saline water for desalination has also been studied, where the rejection rates for the different ions are separately represented for various applied pressures. For seawater, the $MoS_2$ membrane has showed quite promising performance in the study of ion variation.

# 1. Introduction

Water is crucial for sustainable development given that it is the foundation of all life and means of subsistence. Successful management of water will lay the groundwork for achieving many of the 17 Sustainable Development Goals (SDGs), including SDG 6: "Ensure availability and sustainable management of water and sanitation for all." Since the 1980s, the global rate of water use has risen by around 1% annually [1]. The main driver of this constant increase has been the booming demand in emerging and developing nations. Population expansion, socio-economic advancement, and changing purchasing habits all contribute to this growth [2]. With 69% contribution to agriculture from all annual water withdrawals worldwide (including irrigation, livestock, and aquaculture), it is clear that this sector is the biggest water consumer. But water scarcity has been emerging as a pressing problem nowadays. Several factors are responsible for causing such global issue: Population growth, fast urbanization and industrialization have contributed to a rise in water demand, straining already limited freshwater resources. Unsustainable water management methods, ineffective agricultural irrigation techniques, and water source contamination have also increased the depletion and degradation of available water supplies. As a result, fresh water is becoming less accessible, impacting communities, ecosystems, and economies worldwide. With the proven capacity to offer a steady supply of safe drinking water, desalination has established itself as a serious contender in the fight against the global spread of water scarcity. For decades, scientist and researchers have been actively searching for the best water desalination technology that strikes a balance between cost efficiency and energy efficiency.

The development of water desalination processes accelerated in the 1950s and 1960s, a time when oil price was relatively low at around $3 per barrel. Later the oil price index increased to $25/bl [3, 4]. The oil crisis of the early 1970s played a significant role in the growth of the water desalination industry. R.S. Silver licensed the multi-stage flash process in the US and the UK in 1957, and the first multi-stage flash process-based equipment was put in place in 1960 [9] to improve performance and address the shortcomings. ANIL K. RAJVANSH proposed a scheme in 1978 that utilized solar energy to purify sea water for the Indian Thar Deserts [5]. In 1953, desalination plants were installed in Qatar and Kuwait. This marked the expansion of desalination plants worldwide [6]. The growth of desalination technology led to the emergence of various companies involved in the industry. The evaluation conducted by Yusuke Tokui in Kingdom of Saudi Arabia (KSA) compared the sustainability of two kinds of desalination plants: reverse osmosis and multistage flash [7]. While the multistage flash distillation plant had a higher daily water production of 143,600 m3/day,[8] the reverse osmosis plant produced 127,800 m3/day. However, the evaluation revealed that the reverse osmosis plant had lower electricity consumption rates (7.5 kWh/m$^3$) [8] and costs (1.03US\$/m$^3$ )[9] compared to the multistage flash distillation plant (electricity consumption rates 15.84 kWh/m$^3$ and costs 1.33US\$/m$^3$). This suggests that the reverse osmosis process is more energy-efficient and cost-effective in water desalination area.

Reverse osmosis (RO) process indicates a semipermeable membrane-dependent water purification process for removing dissolved minerals, salts, impurities and contaminants out of water. Advanced through the past 40 years, RO membrane technology is contributing to a 44% of the global desalination production facility and an 80% of the desalination plants deployed globally [10]. With better materials and lower costs, membrane desalination has become more popular [10]. Along with reverse osmosis (RO), nanofiltration (NF), and electro dialysis (ED) membrane processes are also available for desalination. NF membranes are a more recent technology of mid-1980s[11]. Nano filtration membranes have been tested for a variation of salt concentrations [12-15]. Even though research indicates that NF cannot, on its own, bring seawater salinity to drinking water level, it has been employed successfully to salinate mildly brackish water as feed. [16, 17]. Seawater can be treated with NF when combined with RO [12, 18, 19]. Initially a strong interest was formed to use cellulose acetate as a semipermeable membrane for reverse osmosis desalination process. Biget had prepared the membrane from aqueous solution containing different metallic perchlorates specially magnesium and beryllium perchlorates and succeed to remove 97% salt ions dissolved in the water [20]. Reid and Spencer had prepared and tested uncoated cellophane (du Pont PT-300), cellulose acetate (du Pont CA-43), polyvinyl alcohol membranes, polyvinyl acetate membranes. They conducted research on the separation of salt from water and found that cellulose acetate demonstrated unique properties compared to other systems they investigated and it possess the capability to remove salt from water [21]. In a study on a pore flow model for reverse osmosis (RO) desalination, Mohammad N. Sarbolouki showed that the size of the membrane pores and their relationship to the size of the transported species are the primary factors in the determination of the significance of bulk flow and diffusional processes.[22]. Several non cellulose membranes also have showed some promising result. T. A. Jadwin had developed a crosslinked hydroxyethyl methacrylate (HEMA) membrane where the crosslinkers were trimethylol propane trimethacrylate (TPT) or ethylene glycol dimethacrylate (EGD) [23].

Usage of membrane materials has led to very promising application in water desalination technology as discussed. Micro and nano porous membranes show significant potentiality in water treatment application because of their variation in geometrical characteristics like: pore size, porosity, tortuosity, pore distribution, surface properties and surface free energy etc. that affects permeability, fouling and filtration capacity. The increasing use of membrane technology pokes to understand the characteristics of synthetic membranes and separation mechanism more in depth. To achieve this specific objective effectively and efficiently, computational techniques like Molecular Dynamics (MD) is applied. The applications of MD simulation in membrane technology are mainly being studied to analyze membrane material development, behavior of ion and salt concentrations: transport phenomena and fouling investigation [24]. In case of membrane material development by MDM (Molecular Dynamics in Membranes) the first hand application of MD simulation in membrane technology was employed to observe polyamide skin layer structure of RO membrane [25]. Future attempts to create high-performance composite RO membranes will involve studying the molecular structure of the membranes' cross-linked poly amide skin layers, which will require the use of molecular simulations and calculations. Hirose et al. proved this in their investigation [30]. In a different study, 250 randomly positioned monomers of the chemicals trimesoyl chloride (also known as TMC) and m-phenylene diamine (MPD) and benzene-1,3,5-tricarboxylic acid chloride (also known as TTC) were used to simulate the polymerization of monomers to create polyamide membranes [26]. Works on polyamide structures of RO membrane in MD has not abundant due to complexity in bond structures and high computation power necessary for doing such. Only a few can be found from the cited works [26-29]. Afterward such difficulties, the research trend was turned out too much simpler nano tube structures like: CNT (Carbon Nano Tube), BNNT (Boron Nitride Nano Tube) etc. to find out the effect of nano technology - their high efficiency in ion rejection and water permeability [30]. The surface chemistry, assembly to one another, and atomic structure of nanomaterials have a substantial impact on their performance [31]. This pulls out the possibility of application to such technologies furthermore. The works of water flow through CNT pores were investigated to find out the entropy, enthalpy and free energy on the surfaces, axial distribution function, density and radial distribution, water binding energy, water occupancy and conduction, chemical potentials etc. phenomena [32-35]. It was also discovered in a different study [36] that a nanotube's size might affect the mobility of water molecules inside the tube. Apart from the nano tubes the usage of graphene has also become very popular to researchers for its lightweight, flexibility and high mechanical strength. Suk and Aluru first gave effort to find the possibility of porous graphene membrane for water transport [37]. Researchers have conducted more research based on the astounding performance of graphene to determine the density of water and constructions, water flux, profile of velocity as well as pressure distribution, salt dismissal, angle distribution function etc. [38, 39]. In recent times, there was also attempt to improvise water permeability of synthetic membranes following biological membranes as benchmark [40]. For instance, the application of aquaporin-1 supports a flow rate of water of 3 molecules/ns along with a low activation energy. [40-42]. The dynamics, functions, and properties of aquaporins have been studied using various MD models, particularly their conduction and selectivity features [43-46].

The focus of the researchers not only was limited to the membrane, but also many studies regarding the behavior of ions and water molecules were also performed [47]. By calculating the coordination numbers of the ions, Hughes and Gale [53] assessed the relationship of ions against a polyamide membrane alongside assessing diffusion coefficients. In another investigation, different NaCl solution concentrations were obtained for various ion diffusion coefficients, together with the radial distribution functions (RDFs) and coordination numbers [48]. In 1996, MD simulations were done focusing on osmosis and reverse osmosis specifically [49]. In this study, researchers examined how altering the solution's concentration or the pressure differential over the membrane affected the transition from FO to RO. Monitoring the dynamic properties and osmotic pressure in addition to recording the solvent and solute density profiles allowed researchers to examine the impact of adsorption close to walls. Thermodynamic concepts that are also being researched can have an impact on the transport features of membranes [50].

The most recent membrane material showing significantly improved performance for desalination application was MoS2 membrane. Application of multifunctional membranes like this are the most advantageous ones. MoS2-based membranes and MoS2 nanosheets both offer several benefits. In comparison to other nanomaterials (Graphene oxide: 207.6 ± 23.4 GPa [51], Metal Organics Framework: 3-7 GPa [52]), MoS2 has a comparatively greater elastic modulus (200-300 GPa [53]), making it easier to prepare water treatment membranes. MoS2, like graphene, can be obtained from natural molybdenite mineral, whilst MOF/COF must be chemically synthesized [54]. Recent research has demonstrated that MoS2 outperforms graphene in many areas, including catalysis, electrochemical characteristics, and others [53, 55, 56]. In one investigation, monolayered MoS2 outperformed graphene, phosphorene, boron nitride, and MoSe2 in terms of water permeability while maintaining an ion rejection rate of >99% [57]. The pervious findings of MoS2 membrane as a desalination membrane motivates to study more about this membrane. From many literatures it was seen emergence to analyze to salt ion effects on this novel membrane material. This research work is mainly focused on this task to find out how different salt solutions in feed water affects salt rejection, water permeability of MoS2 membrane. This work will assist to establish this membrane material further as a practical sea water desalination membrane for synthesis.

## 2. Methodology and System Details

This study makes use of LAMMPS (Large-scale Atomic/Molecular Massively Parallel Simulator), an established molecular dynamics program, to undertake molecular research with a focus on materials analysis [62]. CHARMM-GUI, an effective simulation input generator developed by Lehigh University [58] has been used to model the simulation domain. In order to visualize the result of MD simulation, powerful open visualization and analysis software, OVITO has been used [59]. An input generator CHARMM GUI was used to create a simulation box with a MoS2 monolayer at the center. The MoS2 Nano-membrane divides the 40 x 40 x 140 Å³ simulation box into two compartments; on one side of the membrane, saline water is present, and on the other, pure permeate water. In order to run the simulation, three different saltwater solutions: 1M NaCl, 0.5M MgCl2, and 0.5M CaSO4 aqueous solution—have been added to the feed zone. Initial Concentration of the salts has been taken higher than that of the seawater. For this desalination process, two dimensional MoS2 monolayer has been modeled using 1H unit cell. 3305 no of TIP3P water molecules were present in the domain's feed region, and 1885 no of TIP3P water molecules can be found in the filtered region. Rigid graphene pistons were designed using VMD (Visual Molecular Dynamics). One graphene slab has been placed on the leftmost end of the model, which works as a fixed piston to put an external pressure along +z direction to push saline water across the MoS2 membrane. Another end-graphene slab is placed at the rightmost end of the model which applies 0.1MPa pressure along –z direction to maintain the water density and system stability during simulations. MoS2 membrane remained fixed and periodic boundary is maintained along all direction. For nonbonding interactions, the cutoff distance has been set as 12 Å. Force fields have been applied among the atoms using hybrid potential. Tersoff potential [60] has been employed to increase the accuracy for simulating graphene and Lennard Jones (LJ) parameters have been utilized to describe the interaction with other atoms using mixing rule. Lennard Jones potential parameters of all kinds of atoms are showed in the table 1.

Table1: Lennard Jones potential parameters [61]

| Particles $_{ij}$ | $\varepsilon_{ij}$ (Kcal/mol) | $\sigma_{ij}$ (nm) |
|---|---|---|
| H-H | 0.0460 | 0.40001 |
| O-O | 0.1521 | 3.15057 |
| S-S | 0.3000 | 3.42105 |
| Mo-Mo | 0.0700 | 4.27631 |
| Ca-Ca | 0.1200 | 2.43572 |
| Na-Na | 0.0469 | 2.51367 |
| Cl-Cl | 0.1500 | 4.04468 |
| Mg-Mg | 0.0150 | 2.11143 |

The RO performance of nano porous MoS2 membrane was studied across a wide range of configurational parameters. For three types of pore radius (R=3.39A, 4.2A & 5A), we have examined the effects of the membrane on the desalination performance.

The TIP3P water model, which depicts water molecules as three-point charges (one negative oxygen and two positive hydrogen charges) joined by stiff connections, was utilised in this Non-Equilibrium Molecular Dynamics Model (NEMD). The SHAKE mechanism was implemented to maintain the water molecule rigid for its bond lengths and bond angles. For each simulation, first, the system was energy minimized for 10,000 steps at 0.5 fs. The NVT ensemble is applied during energy minimization to keep the model at 298.15K. Maintaining the temperature during energy minimization helps the system reach a more stable and realistic configuration. Here, the system was configured to the lowest potential energy, allowing the system to reach a stable state before further analysis. After energy minimization, the system undergoes a relaxation phase. Relaxation allows the system to equilibrate and adjust to the minimized energy state. The relaxation step allows the system to settle into a more realistic and representative configuration. The system was equilibrated in the constant number of particles, volume, and temperature (NVT) ensemble for a duration of 1 ns at a temperature of 298.15 K. The NVT ensemble ensures that the system remains at a fixed temperature throughout the equilibration process, allowing for thermal equilibration. The NVT ensemble was used to enable the water molecules to attain their equilibrium density, which was targeted at 1 g/cm^3. The production non-equilibrium simulations were conducted in the NVT ensemble for a duration of 5 ns. To characterize water filtration, a range of external pressures (150 MPa, 250 MPa, and 350 MPa) were applied to a rigid graphene sheet, which was allowed to move. Initially, a pressure of 0.01 MPa was applied to the left piston, while the right piston was subjected to -0.01 MPa, resulting in the squeezing of water between the pistons to fill the pore. This initial step comprised 10,000 simulation steps. Subsequently, the MoS2 atoms were held fixed in space, focusing solely on the water transport and ion rejection properties of MoS2. In the production runs, pressures of 350 MPa, 250 MPa, and 150 MPa were sequentially applied using a piston, while the graphene at the end was maintained at 1 MPa.

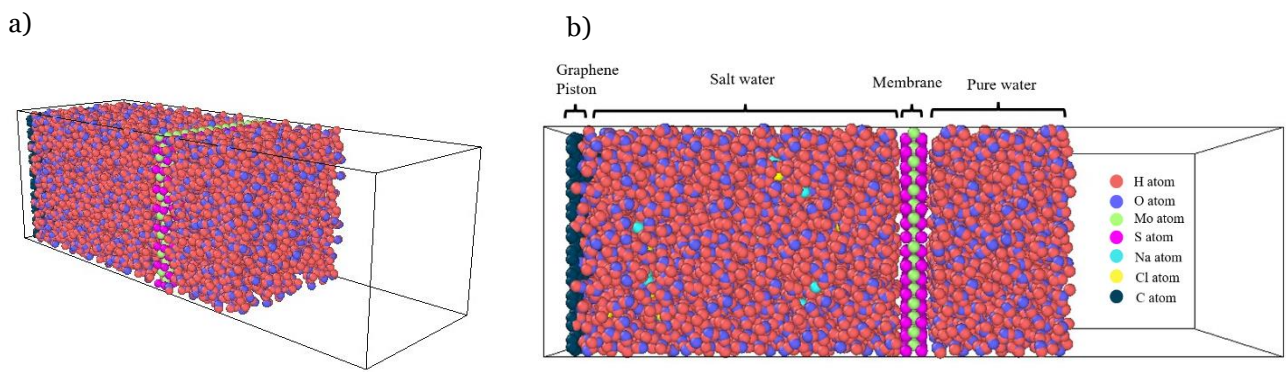

Figure 1: a) Simulation setup and b) grouping of regions

## 3. Results and Discussion

Molybdenum disulfide (MoS$_2$) nanosheets have gained significant attention in recent years as a promising alternative to traditional materials. As a two-dimensional (2D) material, MoS$_2$ nanosheets possess distinctive characteristics in terms of their physicochemical, biological, and mechanical properties. MoS$_2$ has a great potential to be utilized in water-related environmental applications, including membrane-based separation, contaminant adsorption, photocatalysis, sensing, and disinfection etc [53]. Recently, molybdenum disulfide (MoS$_2$) has been explored as a potential membrane for water desalination processes such as- forward osmosis (FO), reverse osmosis (RO), nanofiltration (NF) [62-64].

Numerous studies have been conducted to evaluate the performance of MoS$_2$ in removing sodium chloride (NaCl) from saline water, and the results have been highly promising. In a study by Heiranian et al., it was shown that membranes with pore areas between 20 and 60 Å² achieved an impressive ion rejection rate of over 88% in saline water. The water flux, which measures the rate of water transport, through MoS2 membranes, was significantly higher, ranging from two to five orders of magnitude, in comparison to other nano porous membranes [73]. Zhonglin Cao showed that among various 2D materials with identical pore sizes, single-layer MoS$_2$ consistently outperformed other membranes such as MoSe$_2$, Graphene, BN and Phosphorene. In terms of pressure, even when subjected to an external pressure under 100 MPa, MoS2 demonstrated significantly higher water permeability, outperforming other materials by 20% to 38%, all while maintaining a rejection rate of over 99%. [65]. Ion rejection percentage has been improved from 85% to 98% respectively from monolayer to trilayer of MoS2 membrane [66]. But sea water does not only contain NaCl salt. Though NaCl is the dominant salt in seawater, there are numerous other dissolved salts, including magnesium chloride (MgCl$_2$), calcium sulphate (CaSO$_4$), potassium chloride (KCl) etc. There are 18.9799 grams per kilogram of Cl, 10.5561 grams per kilogram of Na, 1.2720 grams per kilogram of Mg, 0.4001 grams per kilogram of Ca, 0.3800 grams per kilogram of K and 2.6486 grams per kilogram of SO$_4$ and other element are present in sea water [67]. No specific research has been done so far on the performance of MoS$_2$ membranes in rejecting ions other than Na$^+$ and Cl$^-$ ions from saline water. This work focuses on investigating the impact of MoS$_2$ membranes on the removal of salts other than NaCl from saline water during the desalination process. The results presented in our study were obtained through molecular dynamics simulations for 3 cases (1M NaCl water, 0.5M MgCl$_2$ & 0.5M CaSO$_4$). Later number of filtered waters after desalination, water flux through MoS$_2$ membrane and salt rejection percentage has been compared for 3 different salt waters.

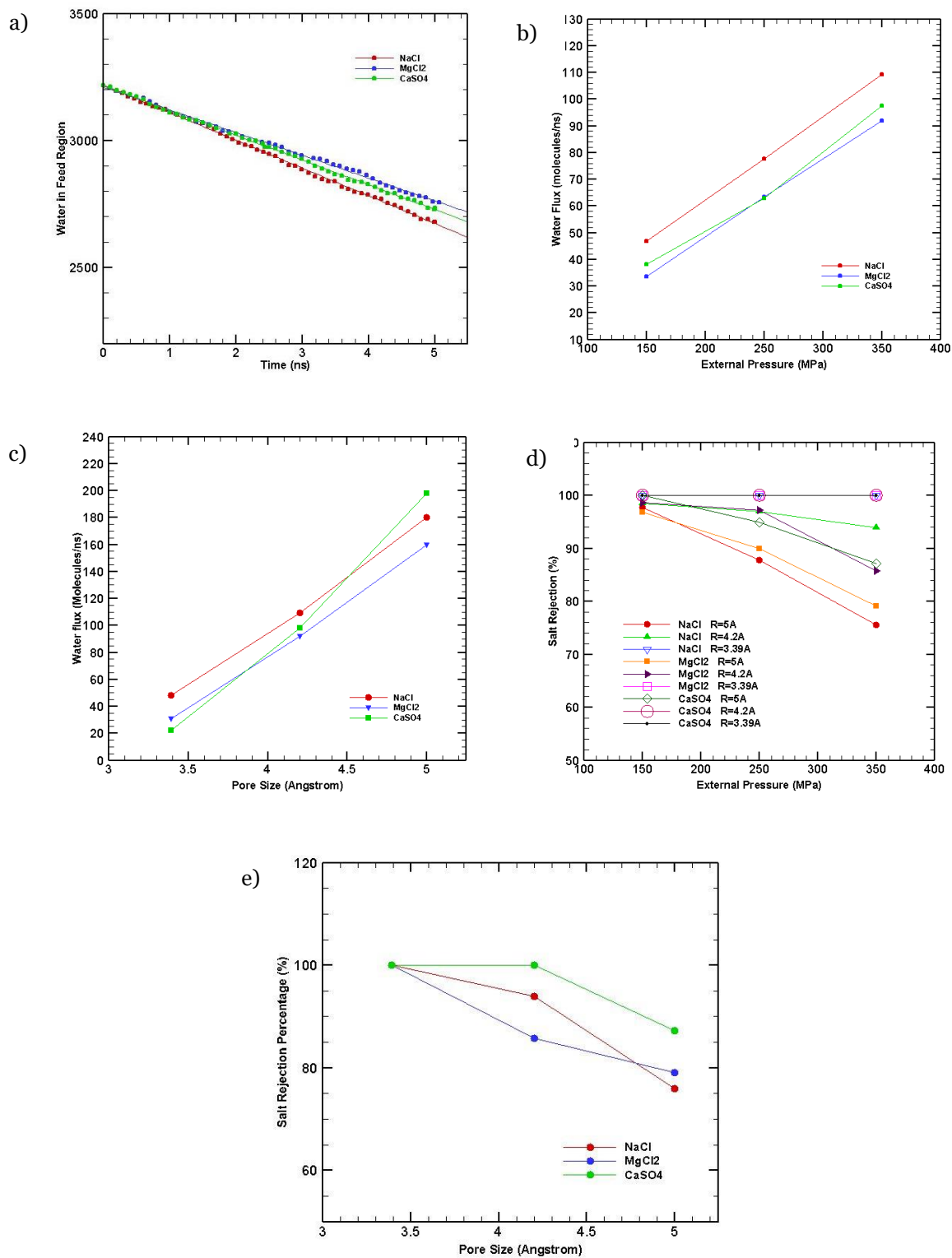

Figure 2: a) Number of water molecules in the feed as a function of simulation time for different salts at a fixed pressure of 350MPa and pore size of 4.2 Å, b) Water flux as a function of external pressure for different salts at a fixed pore size of 4.2 Å, c) The variation of water flux with the variation of pore size for a fixed pressure 350MPa, d) Salt Rejection

Percentage by various pore sizes of different salts as function of external pressure, e) Variation of salt rejection percentage as a function of pore size for a particular external pressure of 350MPa

### *Number of Water in Feed Region:*
In this study, the number of water molecules present in the feed region was calculated and expressed as a function of time in nanoseconds (ns). Additionally, the simulations were performed for three different salts, considering three different pore sizes, and under three distinct pressure conditions. The results of the simulations are presented in the figure, illustrating the changes in the number of water molecules in the feed region over the course of the simulation time period for a particular pressure 350MPa and at a pore size of 4.2 Å. By analyzing these data, we can observe how the presence of different salts can affect the behavior and dynamics of water molecules during water desalination through $MoS_2$ membrane.

The water molecule count in the feed region steadily decreases over time, while the permeate region experiences a simultaneous increase. The consistent water flow rate observed over time suggests a minimal impact of salt concentration on flow rate in the conditions under investigation. From the graph fig 2(a), it is observed that the water molecules in the feed region decreased at a higher rate for NaCl salt solution and the order of water molecule passed is NaCl>CaSO4>MgCl2. In order to track and analyze the behavior of salt ions and water molecules over the course of the simulation, it was necessary to establish a reference time. For each simulation, a specific time of 5ns for the complete simulation period has been defined. A steeper slope on the graph represents a more rapid change in the count of water molecules in the feed region over time, which also signifies the water flux. The slope of the graph also indicates water flux. The water flux is influenced by the type of salts.

### *Water Fluxes:*
One of the objectives of the study was to investigate the variation of water flux with respect to pore sizes and external pressure for three different salts using molecular dynamics simulation. The results from this study have been depicted in the figure 2 (b,c). Water flux is the volume of water passing through the membrane per unit of time. The relation of water flux with pore size can be represented by,

$$Q = \rho * V * A \tag{1}$$

Here, $\rho$ is water density, V represents volumetric flow rate of water or velocity and A indicates the effective cross-sectional area available for water passage. From the equation it can be seen that increasing pore size increases the water flux.

The relation between water flux and pressure can be described by equation: [68]

$$J_W = A(\Delta P - \Delta \pi_m) \tag{2}$$

Where $J_w$ is the volumetric water flux, $A$ is the water permeability coefficient, $\Delta P$ is the applied hydraulic pressure and $\Delta \pi_m$ is the osmotic pressure difference across the active layer. The equation also expresses a proportional relation between water flux and applied external pressure.

As depicted in Figure 2(b, c), the results illustrate that the water flux is related with both external pressure and pore size for all three salts (NaCl, MgCl2, and CaSO4). The larger pore size creates a more substantial effective cross-sectional area, facilitating the passage of water molecules, while increased system pressure enhances water molecule density. These combined effects increase water flux of the desalination system. But the rate of increase is observed quite different, depending on the salt ions present in the solution. The variations are caused by several interatomic interaction effects that are being discussed in the later part.

*Salt Rejection Percentage:*
Another important parameter for measuring the performance of a desalination membrane is salt rejection rate. The salt rejection rate refers to the ability of the membrane to remove salt and other dissolved solids from the feedwater, allowing the production of fresh water. In addition to water permeability, the pore size of nano porous materials also influences salt rejection. Ion rejection can be defined as: [69]

$$(R = 1 - c_{permeate}/c_{feed}) \quad (3)$$

Where, R is salt ion rejection, $c_{permeate}$ denotes the permeate concentration and $c_{feed}$ is the feed concentration. The salt rejection percentage is plotted for various salts and the membrane having different pore sizes, as a function of external pressure in the figure 2(d,e). From the graph, it is evident that there is an inverse relationship for the salts between external pressure and salt rejection rate. As the external pressure decreases, the salt rejection rate increases. This observation suggests that lower pressures enhance the membrane's ability to reject salts more effectively. Even at a relatively high pressure of 350MPa, the salt rejection percentage remains consistently above 80%. In the context of the relationship between pore size and salt rejection rate, it has been observed that as the pore size of desalination membranes increases, the salt rejection rate tends to decrease. For smaller pore sizes (R= 3.39Å), salt rejection percentage is 100%. Furthermore, in comparing different salt solutions, it is noted that CaSO$_4$ demonstrates a higher salt rejection rate compared to MgCl$_2$ and NaCl due to hydration radius and donnan exclusion effect. The discussions are elaborated in details in the part of factors affecting water desalination performance section.

*Permeability Coefficient:*
To measure the water permeability across different pore sizes, permeability coefficient (P) needs to be calculated [70]. The permeability coefficient (P) represents the ability of water molecules to pass through the pores of a given material or membrane. Permeability coefficient, P:

$$p = \frac{J_w}{-V_w \Delta C_s + \frac{V_w}{N_A k_B T} \Delta P} \quad (4)$$

The water flux ($J_w$) is determined by various factors, including the molar volume of water ($V_w$) which is approximately 18.91 ml/mol. Additionally, the concentration gradient of the solute ($\Delta C_s$) plays a role, with a value of 1.0 M. The Avogadro number ($N_A$) and Boltzmann constant ($k_B$) are also involved in the calculations, alongside the temperature (T) set at 300

K. Finally, the applied hydrodynamic pressure (ΔP) in units of MPa further contributes to the determination of the water flux. The permeability coefficients of the NaCl, MgCl$_2$, CaSO$_4$ salts were calculated and presented in table 1.

Table 1: Permeability coefficients of water for the presence of different salts.

| Salts in saline water | Permeability coefficients (water molecule per nano second at 4.2A and 350MPa external pressure |
|---|---|
| NaCl | 41.41 |
| MgCl2 | 34.78 |
| CaSO4 | 36.89 |

***Factors of MoS$_2$ Membrane Affecting Water Desalination:***

The primary factor influencing the transport of water molecules (~2.8 Å van der Waals diameter) through nano porous membranes is the size of the pores [71]. For all three salts (NaCl, MgCl$_2$, and CaSO$_4$), increasing pore size generally leads to higher water permeability. From the outcome of the molecular dynamics simulation of this study, it is evident that increasing pore size results in a decreasing ion rejection rate. So, there is a trade-off between ion rejection rate and permeability when it comes to the pore size of a membrane. For a solute of radius a and the radius of the pore $r_p$, the rejection of solutes in feed water can be approximated using an empirical function based on the ratio of the solute radius to the pore radius, $a/r_p$. For achieving complete solute rejection, it is necessary for the entire pore size distribution to be smaller than the size of the solute of interest ($r_p \leq a$)[69]. In this study, the radius of the pores, $r_p$, is smaller than the radius of the solutes, a used in this study (the radius of the pores, $r_p$ = 3 Å, 4.2 Å, and 5 Å, and the hydrated radius of Na$^+$, Mg$^{2+}$, Ca$^{2+}$, SO$_4$$^{2-}$, and Cl$^-$ are 3.58 Å, 4.28 Å, 3.13 Å, 3.79 Å, and 3.32 Å, respectively) [72, 73].

The transport of solutes can be characterized by considering their size and steric effects, as they often pose significant energy barriers. When solutes attempt to transfer through rigid enclosed pore diameters that are smaller than their own size, it is necessary to reorient the intermolecular bonds and the rearrangement of atoms within the membrane material [74]. The mechanisms underlying the rejection of inorganic salts by nanomembranes can be attributed to the combined effects of size sieving and electrostatic repulsion, which refers to the repulsion force acting between two particles of the same charge. These mechanisms act together to prevent the passage of salts through the membrane [75]. From this study, the salt rejection rate of the MoS$_2$ membrane can be described in the following order: CaSO$_4$ > MgCl$_2$ > NaCl. This aligns with the experimental work of Xu Liang, which shows that MgCl$_2$ has higher salt retention than NaCl [73]. These findings suggest that steric hindrance plays a predominant role in the rejection of monovalent small ions. The observed trend in salt rejection rate can be attributed to the order of hydration radii of the cations, which can be written as Mg2+ > Na+ > Ca$^{2+}$, where the hydrated radii of Na$^+$, Mg$^{2+}$, Ca$^{2+}$, SO$_4$$^{2-}$, and Cl$^-$ are 0.358 nm, 0.428 nm, 0.313nm, 0.379 nm, and 0.332 nm, respectively [72, 73]. But according to the hydration radius, Ca$^{2+}$ is the smallest in size, so the salt rejection was supposed to be the smallest for CaSO$_4$ salt. Here, we see CaSO$_4$ shows the best salt rejection percentage among the other two salts. This happens because of the Donnan exclusion effect, also known as the Gibbs-Donnan effect. This effect refers to the electrostatic repulsion mechanism, in which, due to higher zeta potential and cross-linked degree, electrostatic repulsive force between anions and negatively charged membranes causes high rejection [76]. As the sulfate ion has a higher negative charge than the chloride ion, a higher repulsive force works between the sulfate salt and membrane than that of the chloride salts. This causes a higher salt rejection percentage of $CaSO_4$ than $MgCl_2$ and NaCl. Along with steric hindrance and Donnan repulsive mechanism, an additional repulsion force works between the aqueous solution in the pore and membrane due to the difference in dielectric constants between them. It is believed that, in the case of the salt rejection process, the dielectric mechanism plays a significant role [77].

In an aqueous solution, ions are surrounded by water molecules that become polarized in response to a charged membrane. This polarization creates a layer of water molecules near the ions with charges that oppose the ions themselves. The like charges between the polarized water molecules and the ions result in an additional repulsive force, making it harder for the ions to pass through the membrane. This repulsion, known as dielectric exclusion, adds to the overall electrostatic repulsion between ions and negatively charged membranes, contributing to the rejection of ions in filtration or separation processes. Therefore, in addition to other rejection mechanisms, dielectric exclusion should be well considered in the rejection process. [78]

***Overall Desalination Performance:***

$MoS_2$ membranes have shown promising potential for desalination applications. In this study, several performance parameters (Water flux, Salt rejection percentage and filtered water molecules) have been examined for 3 different pore sizes in $MoS_2$ membrane. Same performance has been investigated for 3 different salt waters (NaCl, $MgCl_2$ and $CaSO_4$). To evaluate the overall desalination performance of MoS2 membranes for three different salts, a comparative analysis is conducted in the figure 3 by comparing the simulation results with the desalination performance of other membranes previously studied in the literature [65, 79]. $MoS_2$ membranes have shown high salt rejection rates for NaCl, with above 90% salt rejection. This indicates that the $MoS_2$ membrane effectively filters out the $Na^+$ and $Cl^-$ ions, resulting in purified water. Additionally, the membrane's ability to remove other salts, such as $MgCl_2$ and $CaSO_4$, suggests its versatility and potential for multi-salt desalination.

$MoS_2$ membranes exhibit higher water fluxes compared to graphene nanopores, with a reported increase of approximately 70%. This suggests that the $MoS_2$ membrane allows for efficient water transport, enabling a higher rate of water permeation through the membrane.

The effective removal of salts other than NaCl, such as $MgCl_2$ and $CaSO_4$, demonstrates the potential of $MoS_2$ membranes for multi-salt desalination. This capability is particularly important when dealing with complex saline water sources that contain various types of salts. The findings highlight the attractive desalination performance of $MoS_2$ membranes, including high salt rejection rates, efficient water flux, and the ability to remove multiple salts. These all have made $MoS_2$ membranes a powerful candidate in the field of water desalination.

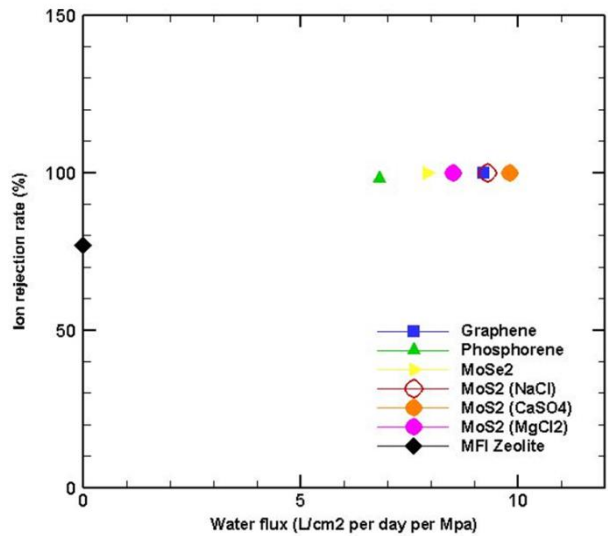

Figure 3: Overall desalination performance of Mos2 membrane for different salt ions.

## 3. Conclusions

Salt rejection and water flux are the two important parameters studied here to examine the performance of $MoS_2$ nano porous membranes in the field of water desalination MoS2 has previously been investigated and proven to be effective in rejecting NaCl salts in saline water. The main objective of this study is to demonstrate that MoS2 is capable of removing other salts found in seawater, in addition to NaCl The results of this study indicate that the MoS2 membrane achieved an above-80% salt rejection rate for MgCl2 and a 100% salt rejection rate for CaSO4. The water flux for both MgCl2 and CaSO4 was found to be similar to that of NaCl.

When compared with other nanomembranes, the overall desalination performance of MoS2 membranes in the presence of different salt ions was superior. Regardless of the type of salt ion, MoS2 outperformed other nanomembranes in terms of salt rejection and water flux. This suggests that MoS2 membranes have broad applicability and effectiveness in removing various salts from water sources. Notably, MoS2 performed exceptionally well in rejecting NaCl, which is the primary salt of concern due to its high concentration in seawater. This finding further underscores the significance and potential of research in this field.

In summary, the study demonstrates that MoS2 nano porous membranes exhibit excellent desalination performance. They can effectively remove a range of salt ions, including NaCl, MgCl2, and CaSO4, while maintaining a high-water flux. The comparative analysis with other nanomembranes indicates that MoS2 surpasses them in desalination efficiency, making it a promising candidate for water desalination applications, particularly in the context of seawater treatment.

**Acknowledgments:**

Authors would like to thank Department of Mechanical Engineering at Bangladesh University of Engineering and Technology (BUET) for providing technical supports for this

*October 12, 2023*

project. This project is also being incentivized by the Research and Innovation Center for Science and Engineering (RISE), BUET. Authors also pay gratitude to Lehigh University for the support with CHARMM-GUI.

*October 12, 2023*

# References

1.  (FAO)., F.a.A.O.o.t.U.N. *AQUASTAT - FAO's Global Information System on Water and Agriculture*. 2022; Available from: https://www.fao.org/land-water/databases-and-software/aquastat/en/.
2.  Nations, U. *UN World Water Development Report 2016*. 2016.
3.  Lior, N., *ADVANCES IN WATER DESALINATION*. 2012, Hoboken, New Jersey: John Wiley & Sons, Inc.
4.  Alnajdi, O., Y. Wu, and J. Kaiser Calautit, *Toward a Sustainable Decentralized Water Supply: Review of Adsorption Desorption Desalination (ADD) and Current Technologies: Saudi Arabia (SA) as a Case Study.* 2020. **12**(4): p. 1111.
5.  Rajvanshi, A.K., *A scheme for large scale desalination of sea water by solar energy*. Solar Energy, 1980. **24**(6): p. 551-560.
6.  Prepared by Desalination Experts Group, O.f.t.W.R.C., *Desalination in the GCC : The History, the Present & the Future, 2014*. 2014, الأمانة العامة: الرياض.
7.  Tokui, Y., H. Moriguchi, and Y. Nishi, *Comprehensive environmental assessment of seawater desalination plants: Multistage flash distillation and reverse osmosis membrane types in Saudi Arabia*. Desalination, 2014. **351**: p. 145-150.
8.  Al-Alshaikh, A. *Seawater desalination in saudi arabia: An overview*. 2012. International Conference on Desalination and Sustainability, Casablanca, 2012.
9.  *Global Water Intelligence: MARKET LEADING ANALYSIS OF PRIVATE WATER PROJECTS*. April 2004, Syed Hossain Publications Son BHD.
10. Greenlee, L.F., et al., *Reverse osmosis desalination: Water sources, technology, and today's challenges*. Water Research, 2009. **43**(9): p. 2317-2348.
11. Singh, R.J.U.w., *A review of membrane technologies: reverse osmosis, nanofiltration, and ultrafiltration*. 1997. **14**(3): p. 21-29.
12. Hilal, N., et al., *Nanofiltration of highly concentrated salt solutions up to seawater salinity*. Desalination, 2005. **184**(1): p. 315-326.
13. Tanninen, J., M. Mänttäri, and M.J.J.o.M.S. Nyström, *Effect of salt mixture concentration on fractionation with NF membranes*. 2006. **283**(1-2): p. 57-64.
14. Wang, D.-X., et al., *Separation performance of a nanofiltration membrane influenced by species and concentration of ions*. 2005. **175**(2): p. 219-225.
15. Wang, D.-X., et al., *Modeling the separation performance of nanofiltration membranes for the mixed salts solution*. 2006. **280**(1-2): p. 734-743.
16. Bohdziewicz, J., M. Bodzek, and E.J.D. Wąsik, *The application of reverse osmosis and nanofiltration to the removal of nitrates from groundwater*. 1999. **121**(2): p. 139-147.
17. Lhassani, A., et al., *Selective demineralization of water by nanofiltration application to the defluorination of brackish water*. 2001. **35**(13): p. 3260-3264.
18. Hamed, O.A.J.D., *Overview of hybrid desalination systems—current status and future prospects*. 2005. **186**(1-3): p. 207-214.
19. Hassan, A., et al., *A new approach to membrane and thermal seawater desalination processes using nanofiltration membranes (Part 1)*. 1998. **118**(1-3): p. 35-51.
20. Huq, H., J.-S. Yang, and J.-W. Yang, *Removal of perchlorate from groundwater by the polyelectrolyte-enhanced ultrafiltration process*. Desalination, 2007. **204**: p. 335-343.
21. Spencer, C.E.R.a.H.G., *ULTRAFILTRATION OF SALT SOLUTIONS AT HIGH PRESSURES* THE JOURNAL OF Physical Chemistry Chemical Physics, 1960. **64**.
22. Sarbolouki, M.N. and I.F.J.D. Miller, *On pore flow models for reverse osmosis desalination*. 1973. **12**(3): p. 343-359.
23. Jadwin, T., A. Hoffman, and W.J.J.o.A.P.S. Vieth, *Crosslinked poly (hydroxyethyl methacrylate) membranes for desalination by reverse osmosis*. 1970. **14**(5): p. 1339-1359.

24. Ebro, H., Y.M. Kim, and J.H. Kim, *Molecular dynamics simulations in membrane-based water treatment processes: A systematic overview*. Journal of Membrane Science, 2013. **438**: p. 112-125.
25. M, H., *The relationship between polymer molecular structure of RO membrane skin layers and their RO performances*. Journal of Membrane Science, 1997. **123**(2): p. 151-156.
26. Harder, E., et al., *Molecular dynamics study of a polymeric reverse osmosis membrane*. J Phys Chem B, 2009. **113**(30): p. 10177-82.
27. Luo, Y., et al., *Computer simulations of water flux and salt permeability of the reverse osmosis FT-30 aromatic polyamide membrane*. Journal of Membrane Science, 2011. **384**: p. 1-9.
28. Shintani, T., et al., *Characterization of methyl-substituted polyamides used for reverse osmosis membranes by positron annihilation lifetime spectroscopy and MD simulation*. 2009. **113**(3): p. 1757-1762.
29. Hughes, Z.E. and J.D. Gale, *Molecular dynamics simulations of the interactions of potential foulant molecules and a reverse osmosis membrane*. Journal of Materials Chemistry, 2012. **22**(1): p. 175-184.
30. Hilder, T.A., D. Gordon, and S.H. Chung, *Salt rejection and water transport through boron nitride nanotubes*. Small, 2009. **5**(19): p. 2183-90.
31. *Riviere, J.C., & Myhra, S. (Eds.). (2009). Handbook of Surface and Interface Analysis: Methods for Problem-Solving, Second Edition (2nd ed.). CRC Press.* [https://doi.org/10.1201/9781420007800](https://doi.org/10.1201/9781420007800).
32. Hummer, G., J.C. Rasaiah, and J.P. Noworyta, *Water conduction through the hydrophobic channel of a carbon nanotube*. Nature, 2001. **414**(6860): p. 188-190.
33. Kalra, A., S. Garde, and G. Hummer, *Osmotic water transport through carbon nanotube membranes*. Proc Natl Acad Sci U S A, 2003. **100**(18): p. 10175-80.
34. Wang, L., R.S. Dumont, and J.M. Dickson, *Nonequilibrium molecular dynamics simulation of water transport through carbon nanotube membranes at low pressurea)*. The Journal of Chemical Physics, 2012. **137**(4).
35. Pascal, T.A., W.A. Goddard, and Y. Jung, *Entropy and the driving force for the filling of carbon nanotubes with water*. 2011. **108**(29): p. 11794-11798.
36. Jia, Y.-x., et al., *Carbon nanotube: Possible candidate for forward osmosis*. Separation and Purification Technology, 2010. **75**(1): p. 55-60.
37. Suk, M.E. and N.R. Aluru, *Water Transport through Ultrathin Graphene*. The Journal of Physical Chemistry Letters, 2010. **1**(10): p. 1590-1594.
38. Cohen-Tanugi, D. and J.C. Grossman, *Water Desalination across Nanoporous Graphene*. Nano Letters, 2012. **12**(7): p. 3602-3608.
39. Hu, G., M. Mao, and S. Ghosal, *Ion transport through a graphene nanopore*. Nanotechnology, 2012. **23**(39): p. 395501.
40. Gong, X., et al., *A charge-driven molecular water pump*. Nature Nanotechnology, 2007. **2**(11): p. 709-712.
41. de Groot, B.L. and H. Grubmüller, *Water permeation across biological membranes: mechanism and dynamics of aquaporin-1 and GlpF*. Science, 2001. **294**(5550): p. 2353-7.
42. Zeidel, M.L., et al., *Reconstitution of functional water channels in liposomes containing purified red cell CHIP28 protein*. Biochemistry, 1992. **31**(33): p. 7436-40.
43. Wang, Y. and E. Tajkhorshid, *Molecular Mechanisms of Conduction and Selectivity in Aquaporin Water Channels*. The Journal of Nutrition, 2007. **137**(6): p. 1509S-1515S.
44. Kumar, M., et al., *Highly permeable polymeric membranes based on the incorporation of the functional water channel protein Aquaporin Z*. Proc Natl Acad Sci U S A, 2007. **104**(52): p. 20719-24.
45. Phillips, J.C., et al., *Scalable molecular dynamics with NAMD*. J Comput Chem, 2005. **26**(16): p. 1781-802.

46. Zhu, F., E. Tajkhorshid, and K. Schulten, *Theory and simulation of water permeation in aquaporin-1*. Biophys J, 2004. **86**(1 Pt 1): p. 50-7.
47. Mi, B., et al., *Physico-chemical characterization of NF/RO membrane active layers by Rutherford backscattering spectrometry*. Journal of Membrane Science, 2006. **282**: p. 71-81.
48. Uchida, H. and M. Matsuoka, *Molecular dynamics simulation of solution structure and dynamics of aqueous sodium chloride solutions from dilute to supersaturated concentration*. Fluid Phase Equilibria, 2004. **219**(1): p. 49-54.
49. Murad, S., *Molecular dynamics simulations of osmosis and reverse osmosis in solutions*. Adsorption, 1996. **2**(1): p. 95-101.
50. Hwang, S.-T., *Nonequilibrium thermodynamics of membrane transport*. 2004. **50**(4): p. 862-870.
51. Suk, J.W., et al., *Mechanical properties of monolayer graphene oxide*. ACS Nano, 2010. **4**(11): p. 6557-64.
52. Zhang, C., et al., *Ultrathin metal/covalent–organic framework membranes towards ultimate separation*. Chemical Society Reviews, 2019. **48**(14): p. 3811-3841.
53. Wang, Z. and B. Mi, *Environmental Applications of 2D Molybdenum Disulfide (MoS2) Nanosheets*. Environmental Science & Technology, 2017. **51**(15): p. 8229-8244.
54. Feng, L., et al., *Unraveling Interaction Mechanisms between Molybdenite and a Dodecane Oil Droplet Using Atomic Force Microscopy*. Langmuir, 2019. **35**(18): p. 6024-6031.
55. Zhang, S., et al., *Photocatalytic wastewater purification with simultaneous hydrogen production using MoS2 QD-decorated hierarchical assembly of ZnIn2S4 on reduced graphene oxide photocatalyst*. Water Research, 2017. **121**: p. 11-19.
56. Xu, Z., et al., *A critical review on the applications and potential risks of emerging MoS2 nanomaterials*. Journal of Hazardous Materials, 2020. **399**: p. 123057.
57. Cao, Z., V. Liu, and A. Barati Farimani, *Why is Single-Layer MoS2 a More Energy Efficient Membrane for Water Desalination?* ACS Energy Letters, 2020. **5**(7): p. 2217-2222.
58. Jo, S., et al., *CHARMM-GUI: A web-based graphical user interface for CHARMM*. Journal of Computational Chemistry, 2008. **29**(11): p. 1859-1865.
59. Stukowski, A., *Visualization and analysis of atomistic simulation data with OVITO–the Open Visualization Tool*. Modelling and Simulation in Materials Science and Engineering, 2010. **18**(1): p. 015012.
60. Rajasekaran, G., R. Kumar, and A. Parashar, *Tersoff potential with improved accuracy for simulating graphene in molecular dynamics environment*. Materials Research Express, 2016. **3**(3): p. 035011.
61. Oviroh, P., et al., *Multilayer Separation Effects on MoS2 Membranes in Water Desalination*. 2021.
62. Li, M.-N., et al., *Forward osmosis membranes modified with laminar MoS2 nanosheet to improve desalination performance and antifouling properties*. Desalination, 2018. **436**: p. 107-113.
63. Xu, G.-R., et al., *Two-dimensional (2D) nanoporous membranes with sub-nanopores in reverse osmosis desalination: Latest developments and future directions*. Desalination, 2019. **451**: p. 18-34.
64. Yang, S. and K. Zhang, *Few-layers MoS2 nanosheets modified thin film composite nanofiltration membranes with improved separation performance*. Journal of Membrane Science, 2020. **595**: p. 117526.
65. Cao, Z., V. Liu, and A.J.A.E.L. Barati Farimani, *Why is single-layer MoS2 a more energy efficient membrane for water desalination?* 2020. **5**(7): p. 2217-2222.
66. Oviroh, P.O., et al., *Nanoporous MoS2 membrane for water desalination: a Molecular Dynamics Study*. 2021. **37**(23): p. 7127-7137.

67. Lyman, J. and R.H.J.J.m.R. Fleming, *Composition of sea water*. 1940. **3**(2): p. 134-146.
68. Baker, R.W., *Membrane technology and applications*. 2012: John Wiley & Sons.
69. Mehta, A. and A.L.J.J.o.m.s. Zydney, *Permeability and selectivity analysis for ultrafiltration membranes*. 2005. **249**(1-2): p. 245-249.
70. Rosenberg, P.A. and A.J.T.J.o.g.p. Finkelstein, *Water permeability of gramicidin A-treated lipid bilayer membranes*. 1978. **72**(3): p. 341-350.
71. Wang, L., et al., *Fundamental transport mechanisms, fabrication and potential applications of nanoporous atomically thin membranes*. 2017. **12**(6): p. 509-522.
72. Liang, X., et al., *Zwitterionic functionalized MoS2 nanosheets for a novel composite membrane with effective salt/dye separation performance*. Journal of Membrane Science, 2019. **573**: p. 270-279.
73. Zuburtikudis, I., et al., *Ionic Liquid functionilized graphene oxide for the adsorption of Ca2+ and Mg2+ ions from saline aqueous feed*. Korean Journal of Chemical Engineering, 2023. **40**.
74. Epsztein, R., et al., *Towards single-species selectivity of membranes with subnanometre pores*. 2020. **15**(6): p. 426-436.
75. Ma, D., et al., *Continuous UiO-66-Type Metal–Organic Framework Thin Film on Polymeric Support for Organic Solvent Nanofiltration*. ACS Applied Materials & Interfaces, 2019. **11**(48): p. 45290-45300.
76. Gumbi, N.N., et al., *Relating the performance of sulfonated thin-film composite nanofiltration membranes to structural properties of macrovoid-free polyethersulfone/sulfonated polysulfone/O-MWCNT supports*. Desalination, 2020. **474**: p. 114176.
77. Bandini, S. and D. Vezzani, *Nanofiltration modeling: the role of dielectric exclusion in membrane characterization*. Chemical Engineering Science, 2003. **58**(15): p. 3303-3326.
78. Yaroshchuk, A.E., *Non-steric mechanisms of nanofiltration: superposition of Donnan and dielectric exclusion*. Separation and Purification Technology, 2001. **22-23**: p. 143-158.
79. Heiranian, M., A.B. Farimani, and N.R.J.N.c. Aluru, *Water desalination with a single-layer MoS2 nanopore*. 2015. **6**(1): p. 8616.